\documentclass[a4paper,10pt]{article}
\usepackage{pos}

\usepackage{graphicx}
\graphicspath{{images/}}

\usepackage[skip=4pt]{caption} 

\title{New physics contributions to $Wtb$ anomalous couplings and top-quark decay}

\author*[a]{Apurba Tiwari}
\author[a]{Sudhir Kumar Gupta}

\affiliation[a]{Aligarh Muslim University,\\
  Aligarh-202002, UP, India}

\emailAdd{atiwari@myamu.ac.in}
\emailAdd{sudhir.ph@amu.ac.in}

\abstract{ In this work, we study the new physics effects arising due the presence of anomalous $Wtb$ vertex through the semileptonic decay 
modes of the top-quark at the Large Hadron Collider. An estimate on the sensitivities of the aforementioned interaction at 5$\sigma$ CL in 
the context of top-quark decay-width measurements and cross-section measurements would also be discussed for the pre-existing 13 TeV 
LHC data and its projections for the proposed LHC runs at 14 TeV, 27 TeV and 100 TeV. We also incorporate the $\mathcal CP$-violating 
effects to such interactions by constructing the $\mathcal CP$-violating asymmetries.
}

\FullConference{%
41st 
International Conference on High Energy physics\\
6-13 July, 2022\\
Bologna, Italy
}

\begin{document}
\maketitle

\section{Introduction}

The charge parity ($\mathcal CP$)-violation phenomenon that was first observed in neutral-kaon decay \cite{Christenson:1964fg} needs 
to be explored beyond the Standard-Model (SM) as SM only provides a small amount of $\mathcal CP$-violation via the CKM-matrices 
\cite{Krawczyk:1987wm} which is not sufficient to explain the matter-antimatter asymmetry of the Universe \cite{Canetti:2012zc}. The 
aim of this present article is to investigate the $\mathcal CP$-violating effects of the anomalous $Wtb$ vertex and provide stringent 
constraints on anomalous $Wtb$ couplings using the measurements of top-quark decay width and cross-section as well as production 
asymmetries. Anomalous interactions in top-quark production and decay have been widely explored in the previous literature 
\cite{Antipin:2008zx}.

We consider an effective field theory approach to parameterise the anomalous $Wtb$ vertex. In this framework, the most general $Wtb$ 
vertex is expressed as \cite{Tiwari:2022nli}:
\begin{eqnarray}
\mathcal L_{Wtb} = -\frac{g}{\sqrt {2}} \bar{b} \left[\gamma^{\mu}\left(C_{1L}P_L + C_{1R} P_R\right)W^{-}_{\mu}
- i\sigma^{\mu\nu}\left(\tilde C_{2L} P_L + \tilde C_{2R} P_R\right) (\partial_{\nu} W^{-}_{\mu})\right]t + h.c.,
\label{eff_lag}
\end{eqnarray}

Where $C_{1L}$, $C_{1R}$, $\tilde C_{2L}~(\frac{C_{2L}}{\Lambda})$ and $\tilde C_{2R}~(\frac{C_{2R}}{\Lambda})$ are dimensionless 
complex anomalous couplings, $P_{L,R} = \frac{1}{2} (1 \mp \gamma_5)$ and $\Lambda$ is the energy scale. In SM at tree level $C_{1L}$ = 
$V_{tb}$ = 1 and other couplings are zero.

\section{Numerical Analysis}

We begin the analysis by incorporating the Lagrangian given in Eq. \ref{eff_lag} into $\tt Feynrules$ \cite{Christensen:2008py}, which 
then interfaced with $\tt Feyncalc$ \cite{Shtabovenko:2016sxi} for further simulations. The decay level $\mathcal CP$-asymmetries will 
be defined as:
\begin{eqnarray}
{\cal A}^{\Gamma}_{SM} &=& \frac{\Delta \Gamma_{t\to b W}}{\Gamma_{t\to b W}} \simeq \frac{{\rm Im} \left(|{\cal M}|^2_{t\to b W}\right) }
{{\rm Re} \left(|{\cal M}|^2_{t\to b W}\right)}
\label{as_gamx}
\end{eqnarray}
where $|{\cal M}|^2_{t\to b W}$ is the matrix-element squared for the process of top (anti-top) decay into $b~(\bar{b})$-quark and 
$W^{+}~(W^{-})$-boson. The expression for the relative decay width of the top-quark with anomalous coupling to the SM decay width is:
\begin{eqnarray}
R^{\Gamma} &=& \frac{\Gamma_{t\to b W}}{\Gamma^{SM}_{t\to b W}} = 1 - \frac{M_W}{(1 + 2 \eta^2)}[6 \eta C_R - M_W(\eta^2 + 2)(C_L^2 + C_R^2)]
\label{Rgam}
\end{eqnarray}
where $\eta = \frac{M_W}{m_t}$, $C_L = |C_L| e^{i\theta}$ and $C_R = |C_R| e^{i\phi}$.

\begin{figure*}[h!]
\begin{tabular}{c c}
\hspace{-0.3cm}
\includegraphics[width=\textwidth,height=5.15cm]{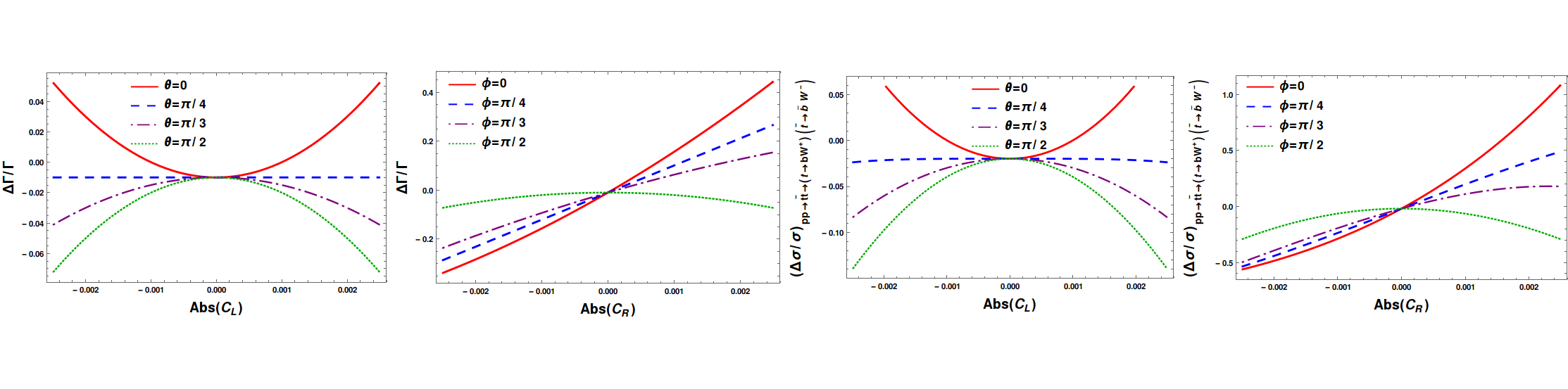}\hspace{-0.3cm}
\end{tabular}
\caption{Dependence of decay width of top-quark and cross-section on the moduli of anomalous couplings for different values of $\theta$ and 
$\phi$ for the cases with $\left|C_R\right|$ = 0 and $\left|C_L\right|$ = 0.
}
 \label{width_xsec_plots}
\end{figure*}

In Fig. \ref{width_xsec_plots}, we show the dependence of the relative change in decay width and cross-section, $\frac{\Delta 
\sigma}{\sigma}$ on moduli of the anomalous coupling at different values of phases $\theta$ and $\phi$. We observe that the decay 
width as well as cross-section is more sensitive to coupling $C_R$ and the contribution from the coupling $C_L$ is negligible. In 
Table \ref{width_xsec_bounds}, we present the constraints on anomalous couplings $C_L$ and $C_R$ at 2.5$\sigma$ C.L. (when only one 
anomalous coupling is taken non zero at a time) obtained from the top-quark decay width measurements.
\begin{table}[h!]
\centering
\scalebox{0.8}{
\renewcommand{\arraystretch}{2.0}
\begin{tabular} {c|c|c}
\hline
                                                                        & $C_L~(\times 10^{-3})$     & $C_R~(\times 10^{-3})$ \\
\hline\hline
$\left(\frac{\Delta \Gamma}{\Gamma}\right)_{t \to bW}$                  & $ -5.86 \le C_L \le 5.86 $ & $ -1.84 \le C_R \le 1.95 $\\
$\left(\frac{\Delta \sigma}{\sigma}\right)_{pp \to t\bar{t}}^{13 TeV}$  & $ -2.62 \le C_L \le 2.62 $ & $ -0.40 \le C_R \le 0.40 $\\
\hline\hline
\end{tabular}}
\caption{Bounds on anomalous couplings $C_L$ (when $C_R$ = 0) and $C_R$ (when $C_L$ = 0) at 2.5$\sigma$ C.L. obtained from measurements of 
top-quark decay width and top-pair production cross-section at the LHC with $\sqrt{s}$ = 13 TeV.
}
\label{width_xsec_bounds}
\end{table}

In the same way as the decay width, the production asymmetries could be estimated using the formula,
\begin{eqnarray}
{\cal A}^{\sigma}_{SM} &=& \frac{\Delta \sigma_{p p \longrightarrow t (\to b W^+) {\bar t} (\to {\bar b} W^-)}}
{\sigma_{p p \longrightarrow t (\to b W^+) {\bar t} (\to {\bar b} W^-)}} \simeq \left(\frac{{\rm Im} \left(|{\cal M}|^2_{t\to b W}\right)}
{{\rm Re} \left(|{\cal M}|^2_{t\to b W}\right)}\right)^2.
\label{as_sigx}
\end{eqnarray}

\begin{figure*}[h!]
\begin{tabular}{c c}
\hspace{-0.3cm}
\includegraphics[width=\textwidth,height=5.15cm]{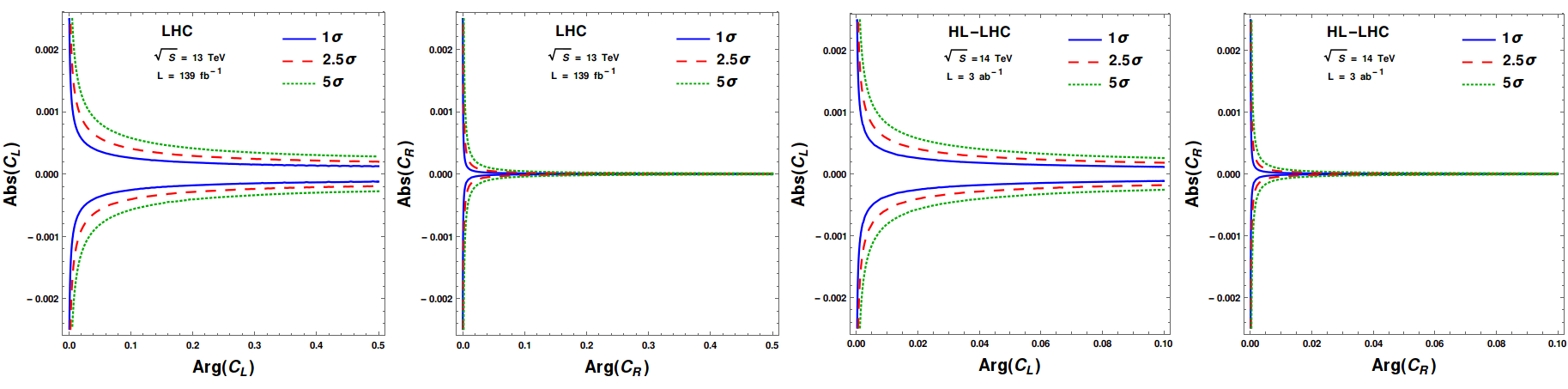}\hspace{-0.3cm}\\
\includegraphics[width=\textwidth,height=5.15cm]{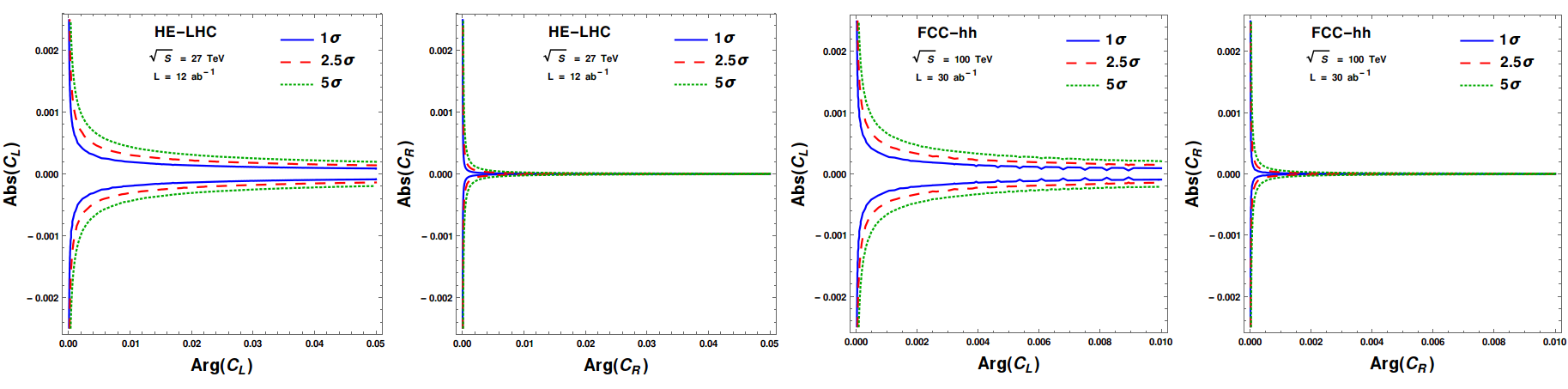}\hspace{-0.3cm}
\end{tabular}
\caption{1$\sigma$, 2.5$\sigma$ and 5$\sigma$ C.L. regions in the Abs($C_L$)-Arg($C_L$) plane and Abs$(C_R)$-Arg($C_R$) plane allowed by the 
production asymmetry at LHC with $\sqrt{s}$ = 13 TeV, HL-LHC with $\sqrt{s}$ = 14 TeV, HE-LHC with $\sqrt{s}$ = 27 TeV and FCC-hh with 
$\sqrt{s}$ = 100 TeV.
}
 \label{asymm_plots}
\end{figure*}

Fig. \ref{asymm_plots} presents the 1$\sigma$, 2.5$\sigma$ and 5$\sigma$ regions in Abs($C_L$)-Arg($C_L$) plane and 
Abs($C_R$)-Arg($C_R$) plane allowed by the production asymmetries at LHC with $\sqrt{s}$ = 13 TeV, HL-LHC with $\sqrt{s}$ = 14 TeV, 
HE-LHC with $\sqrt{s}$ = 27 TeV and FCC-hh with $\sqrt{s}$ =100 TeV. It should be noted that an approximate prediction of the 
constraints on the phase and moduli of the anomalous couplings $C_L$ and $C_R$ at 2.5$\sigma$ C.L. can be given from Fig. 
\ref{asymm_plots}, however an exact calculation has been carried out to obtain the limits and the values obtained are presented in 
Table \ref{asymm_bounds}.

\section{Conclusions}

In this study, we have explored the $\mathcal CP$-violating effects of the anomalous $Wtb$ vertex in the context of top-quark via its 
decay into a b-quark and a W-boson. An estimate of the constraints on the anomalous couplings $C_L$ and $C_R$ have been presented 
using top-quark decay width and cross-section measurements as well as production asymmetries. The values of the limits obtained on the 
couplings $C_L$ and $C_R$ are presented in Tables \ref{width_xsec_bounds} and \ref{asymm_bounds}.
\begin{table}[h!]
  \begin{center}
\scalebox{0.8}{
  \renewcommand{\arraystretch}{1.6}
    \begin{tabular}{c|c|c|c}
Collider & $\sqrt{s}$, $\mathcal \int Ldt$ & $\left|C_L\right|~(\times 10^{-4})$ & $\left|C_R\right|~(\times 10^{-4})$ \\
\hline\hline
LHC      & 13 TeV, 139 fb$^{-1}$           &  1.82  & 0.03 \\
\hline
HL-LHC   & 14 TeV, 3.0 ab$^{-1}$           &  0.81  & 0.006 \\
\hline
HE-LHC   & 27 TeV, 12.0 ab$^{-1}$          &  0.44  & 0.0017 \\
\hline
FCC-hh   & 100 TeV, 30.0 ab$^{-1}$         &  0.21  & 0.0004 \\
\hline\hline
    \end{tabular}}
\caption{Bounds on the moduli of the anomalous couplings $C_L$ (when $C_R$ = 0) and $C_R$ (when $C_L$ = 0) for CP-violating phase, 
$\theta=\phi=\frac{\pi}{4}$ at 2.5$\sigma$ C.L. obtained from production asymmetries at LHC, HL-LHC, HE-LHC and FCC-hh.
}
 \label{asymm_bounds}
 \end{center}
\end{table}

\end{document}